\documentclass[aps,preprint,showpacs,nofootinbib,12pt]{revtex4}
\usepackage{graphicx}
\usepackage{epsfig}
\usepackage[dvips]{color}

\begin{document}

\title{ Fermion correction to the mass of the scalar glueball in QCD sum rule }

\author{ Xu-Hao Yuan $^{1}$   \footnote{segoat@mail.nankai.edu.cn},
         Liang Tang $^{1}$  \footnote{tangliang@mail.nankai.edu.cn}  }

\affiliation{
  $^{1}$ School of Physics, Nankai University, Tianjin 300071, China }

\begin{abstract}
Contributions of fermions to the mass of the scalar glueball
$0^{++}$ are calculated at two-loop level in the framework of QCD
sum rules. It obviously changes the coefficients in the operator
product expansion (OPE) and  shifts the mass of glueball.
\end{abstract}
\maketitle

\section{introduction}

Quantum Chromo-dynamics (QCD) predicts the existence of glueballs.
After a long time of experimental and theoretical exploration for
glueballs, there is no obvious evidence to confirm its existence
yet, even though people find several glueball candidates, such as
$f_0(1710)$ and $\eta(1405)$\cite{Zheng:2009a24}. There are also
predictions on the mass of glueballs in various theoretical
frameworks. Among all the theoretical approaches, the estimation on
the mass of glueballs by the Lattice QCD and QCD sum rules, seems to
be closer to reality. The Lattice QCD predicts the mass of the
scalar glueball ($0^{++}$) as
$1.5\sim1.8$GeV
\cite{Bali:1993fb,Chen:1994uw,Morningstar:1999rf,Vaccarino:1999ku,
Liu:2000ce,Liu:2001je,Ishii:2001zq,Loan:2005ff,Chen:2005mg}.
With the QCD sum rules, Novikov  and Narison evaluated the mass of
scalar glueball as $700-900
\mathrm{MeV}$\cite{Novikov:1980npb165,Pascual:1982plb,Narison:1984zpc},
whereas Bagan and Steele considered the radiative corrections and
obtained the mass as $1.7\mathrm{GeV}$\cite{Bagan:1990plb}. Later,
based on Bagan and Steele's work, Huang {\it et al.}, re-estimated
the mass and found a small shift to $\sim
1.66\mathrm{GeV}$\cite{Huang:1999prd}. The difference is so large
that one has reason to doubt if there indeed exists theoretical
discrepancy. One compelling motivation is that one needs to make a
complete calculation which should add up the contributions which
were neglected in previous calculations. That is the aim of this
work.

We have repeated Bagan and Steele's derivations, and noticed that
they neglected the contributions of the fermions  to the radiative
corrections. Namely, they neglected contributions of the loops
involving fermion propagators and quark condensates by setting $C_f$
which is a coefficient related to quark flavors, to be zero in
\cite{Bagan:1990plb}. They argued, such contributions were small
compared to others. In this work we include the contributions of the
loops containing fermion propagators and quark condensates to the
correlation function $\Pi(q^2)$. With this correction, we set a
proper platform at $s_0=3.8$ GeV, where $s_0$ represents the
threshold for the continuum states, and eventually we determine the
mass of the glueball as $1721~\mathrm{MeV}$. This value is
compatible with that obtained by Bagan and Steele, a bit larger than
that Huang et al. achieved, but within a tolerable error region, all
of them are consistent with each other.

\section{scalar glueball QCD sum rule}

The correlation function for scalar glueballs is defined as:
\begin{eqnarray}\label{cf defination}
 \Pi(q^2)=i \int d^4 x e^{iq\cdot x} \langle
 0|T\{j(x)j(0)\}|0\rangle,
\end{eqnarray}
where, $ j(x)=\alpha_s G_{\mu\nu}^a (x) G^{a \mu\nu}(x) $  stands
for the current of the $ 0^{++} $ glueball. By the operator product
expansion (OPE), the correlation function can be further written as:
\begin{eqnarray}\label{Pi defination}
 \Pi(q^2)&=&\sum_n C_n(q^2) \langle0|\hat{O}_n|0\rangle\nonumber\\
 &=&[a_0+a_1 \log\frac{Q^2}{\nu^2}](Q^2)^2 \log\frac{Q^2}{\nu^2}+[b_0+b_1
 \log\frac{Q^2}{\nu^2}]\langle \alpha_s G^2 \rangle+[c_0\nonumber\\
 &+&c_1\log\frac{Q^2}{\nu^2}]\frac{\langle g_s G^3 \rangle}{Q^2}
 +\frac{d_0}{Q^2}^2\langle \alpha_s^2G^4 \rangle+...
\emph{~~;~~~~~~}Q^2\equiv-q^2>0,
\end{eqnarray}
where, $C_n (q^2)$ are the Wilson coefficients, and those operators
$\hat{O}_n$ have already well defined in
ref.\cite{Colangelo:2000hep-ph}. A more convenient function form
$\mathcal{R}_k(\tau,s_0)$ may be used in later calculations:
\begin{eqnarray}\label{R defination}
\mathcal{R}_k(\tau,s_0)&\equiv&\frac{1}{\tau}\hat{L}[(Q^2)^k
\Pi(-Q^2)]-\frac{1}{\pi}\int^\infty_{s_0}d s s^k e^{-s\tau} Im
\Pi(s)\nonumber\\
&=&\frac{1}{\pi}\int^{s_0}_0d s~s^k e^{-s\tau} Im \Pi(s),
\end{eqnarray}
where, $\hat{L}$ the Borel transformation, $\tau$ is the Borel
parameter, and $s_0$ represents the threshold for the continuum
states. substituting Eq (\ref{Pi defination}) into Eq (\ref{R
defination}), then we have\cite{Bagan:1990plb} (for $k\geq-1$):
\begin{eqnarray}
 \mathcal{R}_{-1}(\tau,s_0)&=&-\frac{a_0}{\tau^2}[1-\rho_1(s_0\tau)]
 +2\frac{a_1}{\tau^2}\{\gamma_E+E_1(s_0\tau)+\log s_0\tau+e^{-s_0\tau}\nonumber\\
 &-&1-[1-\rho_1(s_0\tau)]\log\frac{s_0}{\nu^2}\}+\Pi(0)-\{b_0-b_1[\gamma_E+\log\tau\nu^2\nonumber\\
 &+&E_1(s_0\tau)]\}\langle \alpha_s G^2 \rangle-
 \{c_0+c_1[1-\gamma_E-\log\tau\nu^2-E_1(s_0\tau)
 \label{R-1 formation}\\
 &+&\frac{e^{-s_0\tau}}{s_0\tau}]\}
 \langle g_sG^3 \rangle-\frac{d_0}{2}\langle \alpha_s^2G^4
 \tau^2\rangle\nonumber\\
 \mathcal{R}_k(\tau,s_0)&=&(-\frac{\partial}{\partial\tau})^{k+1}\mathcal{R}_{-1}(\tau,s_0)
 \label{R-k defination}\\
 \mathrm{where},&~&\rho_k(x)\equiv e^{-x}\sum^k_{j=0}x^j/j!,~~
 E_1(x)\equiv\int^\infty_xdye^{-y}/y,\\\label{useful function defination}
 \mathrm{and},&~&\gamma_E=\mathrm{Euler's~
 constant}\approx0.5772.
\end{eqnarray}
With the function $\mathcal{R}_k(\tau,s_0)$, the mass of the scalar
glueball is:
\begin{eqnarray}\label{mass formation}
 \centering
 M_{gg}^2(\tau,s_0)=-{\partial\over\partial\tau}(\log{\mathcal{R}_k})
  =\frac{\mathcal{R}_{k+1}(\tau,s_0)}{\mathcal{R}_k(\tau,s_0)}.
\end{eqnarray}
Although the mass of glueball is determined by a sum of
$\mathcal{R}_k$ with different integer k values, only
$\mathcal{R}_1$ is mostly important and kept in the final result.
The reason is that only $\mathcal{R}_1$ is reliable for
determination of the $0^{++}$ glueball mass\cite{Huang:1999prd}.
According to this comment above, for the $\mathcal{R}$ with $k>0$,
only $a_i$ ($i=0,1$), $b_1$ and $c_1$ in (\ref{R-1 formation}) can
affect the final mass of the glueball, since other coefficients will
vanish through the derivation in Eq(\ref{R-k defination}).

\section{quark contribution in correlation function $\Pi(q^2)$}

\begin{figure}\label{fig-quarkloop}
 \centering
  \includegraphics
   {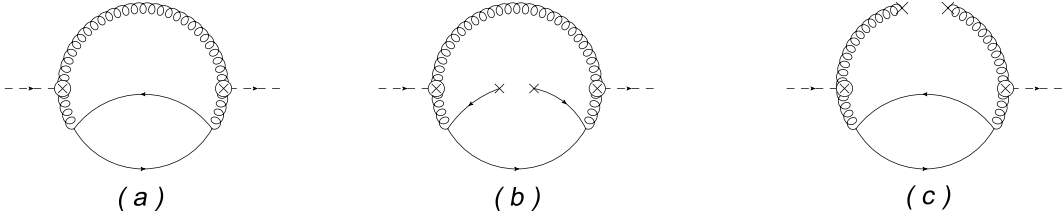}
    \begin{quote}
     \caption[]{(a) the pertubative part; (b) the quark condensation part; (c) the gluon condensation part}
    \end{quote}
\end{figure}

With the fermion contributions containing the fermion propagators
and the quark condensations, the correlation function $\Pi(q^2)$ can
be divided as:
\begin{eqnarray}\label{defination of the Pi's division}
 \Pi(q^2)&=&\Pi^1(q^2)+\Pi^f(q^2)
\end{eqnarray}
where, the $\Pi^1(q^2)$ is the correlation function without fermion
contributions, and was given by Bagan and
Steele\cite{Bagan:1990plb}. The $\Pi^f(q^2)$ stands for the fermion
part. With OPE, we have
\begin{eqnarray}
\Pi^f(q^2)&=&C^f_0\langle\hat{O}_0\rangle+C^f_3\langle\hat{O}_3\rangle+C^f_4\langle\hat{O}_4\rangle+...,
\end{eqnarray}
where, $\hat{O}_n$ is defined above. $C^f_0$ is the Wilson coefficient for
the unit operator $\hat{O}_0$ in Fig {1-a}, and in
$\overline{\emph{MS}}$ scheme, we have:
\begin{eqnarray}\label{c0 defination}
 C^f_0(q^2)&=&({\alpha_s\over\pi})^3(Q^2)^2(5\log{Q^2\over\nu^2}-2\log^2{Q^2\over\nu^2}).
\end{eqnarray}

$C^f_3$ is the Wilson coefficient for operators with quark
condensate. From Fig(1-b), we have:
\begin{eqnarray}\label{c3 definition}
 C^f_3(q^2)\langle\hat{O}_3\rangle&=&
  {\alpha_s\over\pi}
   [
    {56\over9}m_q^3
     (
      E_1(m_q^2\tau)-E_1(s_0\tau)+\log{m_q^2}
       (
        1-\rho_0(0)
       )
     )
    +{8m_q\over\tau}
     (
      \rho_0(\tau s_0)\nonumber\\
      &-&\rho_0(m_q^2\tau)
     )
    +{80\over9\tau^3m_q^3}
     (
      \rho_2(m_q^2\tau)-1
     )
   ]
  \langle q\bar{q}\rangle
  +{\alpha_s\over\pi}
   [
    {10m_q\over9}
    (
     -E_1(m_q^2\tau)+E_1(\tau s_0)\nonumber\\&-&\log{m_q^2}
      (
       1-\rho_0(0)
      )
     )
   +{2\over\tau^2m_q^3}
    (
     \rho_1(m_q^2\tau)-1
    )
   +{40\over9m_q^5\tau^3}
    (
     1-\rho_2(m_q^2\tau)
    )
   ]\nonumber\\
  &\times&\langle g_sq\bar{q}G\rangle
  +{\alpha_s^2}
   [
     {32\over243}
     (
      -
       (
        E_1(m_q^2\tau)-E_1(\tau s_0)
       )
      -\log{m_q^2}
       (
        1-\rho_0(0)
       )
      )\nonumber\\
    &+&{32\over27m_q^4\tau^2}
     (
      1-\rho_1(m_q^2\tau)
     )
    +{1024\over243m_q^6\tau^3}
     (
      \rho_2(m_q^2\tau)-1
     )
   ]\langle q\bar{q}q\bar{q}\rangle,
\end{eqnarray}
where, $\rho_k(x)$ and $E_1(x)$ are defined in Eq (\ref{useful
function defination}), and  $q$ stands for the u, d and s quarks.
Later, we will show that, the contributions from the parts with
quark condensates are less than 1\%, so in general, can be safely
ignored.

\begin{figure}\label{fig-mass}
 \centering
  \includegraphics
   {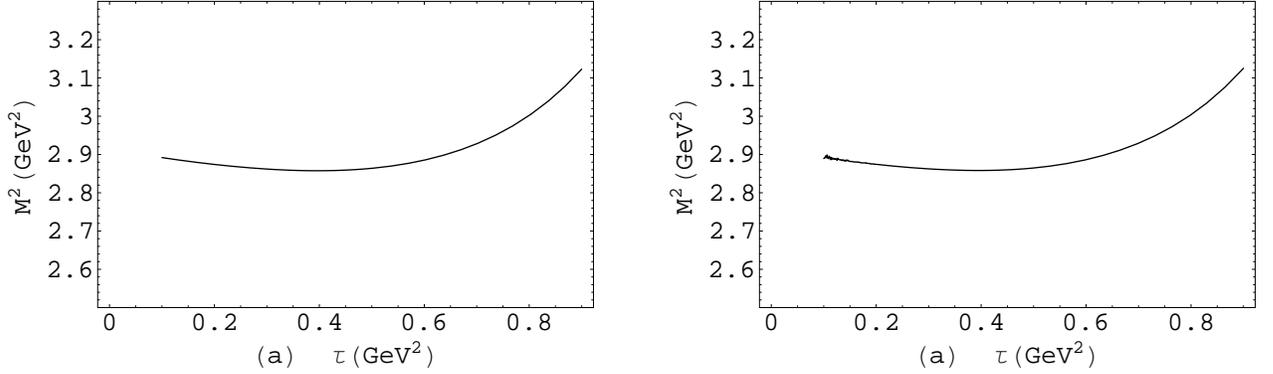}
    \begin{quote}
     \caption[]{(a) the mass without $C^f_3$ at $s_0=3.8\emph{GeV}^2$; (b) the mass with $C^f_3$ at $s_0=3.8\emph{GeV}^2$}
    \end{quote}
\end{figure}

$C^f_4(q^2)$ is the Wilson coefficient for the operators with gluon
condensates. Generally, there are two ways to calculate the Wilson
coefficients\cite{reinders:1985pr}, one  is the plane wave method,
by which $\Pi^1(q^2)$ is given (See Eq (\ref{defination of the Pi's
division})). The another is the fixed-point gauge technique, by
which we have:
\begin{eqnarray}\label{c4 difination}
 C^f_4(q^2)\langle\hat{O}_4\rangle
  &=&
   {\alpha_s^2\over\pi}
    (
     -{13\over3}
     +2\log{-q^2\over\nu^2}
    )
    \langle\alpha_sG^2\rangle.
\end{eqnarray}
Substituting all the corrections ((\ref{c0 defination}), (\ref{c3
definition}), and (\ref{c4 difination})) back into Eq
(\ref{defination of the Pi's division}), we have the correlation
function $\Pi(q^2)$ and the various functions $\mathcal{R}_k$ with
corrections from fermions. The result is shown in fig(2):
We choose the reasonable platform at $\tau\in\{0.4,0.6\}$ in the
region $3.6~\mathrm{GeV}^2<s_0<4.2~\mathrm{GeV}^2$. Within the
platform, the ratio of the contribution of the unit operator term
$C_0(q^2)\hat{O}_0$, which stands for the pertubative part, to the
mass determined by $\mathcal{R}_k$ is more than $90\%$, it enables
the OPE expansion to converge sufficiently fast. Besides, in the
region, the ratio of the contribution of the continuum part to the
mass is less than $30\%-40\%$. It implies that this value of $s_0$
is appropriate for the quark-hardron duality. Within this platform
we have obtained the $0^{++}$ glueball mass as:
$1.721\pm0.065~\mathrm{GeV}^2$, where the error is caused by the
variable of the $s_0$ in the region, meanwhile, the error caused by
the variable of the Borel parameter $\tau$ is very tiny so that it
can be ignored. Fig (2-a) and Fig (2-b) show that, the quark
condensate contributes little to the mass of the glueball, since the
$C^f_3$ given in Eq (\ref{c3 definition}) turns to zero at
$mq\rightarrow 0$. That is why we could directly determine the mass
of glueball by neglecting $C^f_3$ as long as the mass of light
quarks is small.

\section{Conclusion and discussion}
In this paper, we analyze the contribution of the diagrams involving
internal fermion lines and quark condensates. Following the
traditional way, we determinate the mass of the $0^{++}$ scalar
glueball as $1.728\pm0.132~\mathrm{GeV}^2$. Comparing the result of
Huang ($m\sim1.66\mathrm{GeV}^2$), a little shift of the mass is
resulted in by taking the fermion condensation into accout.

\section*{Acknowledgement}
This paper is completed under direction of Profs. Xue-Qian Li and
Mao-Zhi Yang. This work is supported by the National Natural Science
Foundation of China (NNSFC) and the Special Grant for the Ph.D
program of the Education Ministry of China.\\

\end{document}